\documentclass{osa-article}

\journal{osajournal}

\articletype{Research Article}

\begin{document}

\title{A Statistical Model for Imaging Systems}

\author{Jianfeng Zhou\authormark{1,2,*}}

\address{\authormark{1}Center for Astrophysics, Tsinghua University, Beijing 100084, China\\
\authormark{2}Xingfan Information Technology Co., Ltd.(Ningbo), Zhejiang 315500, China
}

\email{\authormark{*}zhoujf@tsinghua.edu.cn} 



\begin{abstract}
The behavior of photons is controlled by quantum mechanics, not as deterministic as classical optics shows. To this end, we defined a new statistic $Z$, which is equal to the variance minus the expectation or mean. Then, we established a statistical model for imaging systems and obtained three fundamental imaging formulas. Among them, the first formula is entirely consistent with the classic convolution equation. The second and third ones link the $Z$ quantities of the object and noise elegantly with the $Z$ image and covariance image, revealing new laws. Consequently, besides the flux density, the $Z$ quantity of an object is also imageable, which opens a new realm for imaging systems to explore the physical world.
\end{abstract}

\section{Introduction}
\label{sec:introduction}
In classical optics, the behavior of photons can be described by deterministic waves \cite{born2013principles}, while the actual situation is that quantum mechanics should depict the act of photons. Therefore, We need a statistical model based on quantum mechanics to describe an imaging system completely.

For a shift-invariant imaging system, in discrete cases,
the classical optics would represent the system by the following convolution equation \cite{jonathan2005digital}
\begin{equation}
I(k) = \sum_{j=-J}^J O(k-j) p(j) + N(k), 
\label{eq:convolution}
\end{equation}
where $O$ is the object, $p$ is the Point Spread Function(PSF), $I$ denotes the observed image, $N$ denotes the noise. And, $k-j$ represents a position on the object plane, and $k$ represents a position on the image plane.

A more general imaging system needs a modulation equation for representation, which is shown in
\begin{equation}
I(k) = \sum_{j=-J}^J K(k,j) O(j) + N(k), 
\label{eq:modulation}
\end{equation}
where the shift variant kernel $K(k,j)$ describes the properties of the imaging system.

Quantum optics originated from the photon correlation experiments completed by Hanbury-Brow and Twiss in the 1950s \cite{brown1956correlation,brown1956test}. Later, Mandel proposed a semi-classical theory to explain the photon correlation observed from conventional sources of light \cite{mandel1959fluctuations,mandel1963progress}. Glauber developed a fully quantum mechanical theory of photons in 1963 \cite{glauber1963quantum,glauber1963coherent}. Quantum optics points out that in addition to the classical correlations, photons also have anti-correlations, corresponding to the photon's bunching and anti-bunching effects \cite{teich1988photon}.

In this paper, we use some actual results of quantum optics to build a statistical model to describe an imaging system fully. In this model, we treat the behavior of photons as a random variable and the imaging process as a stochastic process controlled by the imaging device \cite{goodman2000statoptics}. After that, we use statistical tools to derive the interdependence among the statistical features of the objects, images, and noise.

In Section 2 of this paper, we describe the four essential elements of an imaging system, i.e., the object, the imaging device, the noise, and the observed images. In particular, we define a new but reasonably simple statistic $Z$, which is equal to the variance minus the expectation or mean. In Section 3, we establish the statistical model for imaging systems, present the fundamental joint probability mass function, and derive the expressions of the mean image, variance image, $Z$ image, and covariance image. In Section 4, we present some numerical simulations which demonstrate the inferences of the model. The final parts of the paper are the discussion and the conclusions.

\section{Four Elements of an Imaging System}
\label{sec:fourComponents}
To describe an imaging system in the traditional sense, what we need is the point spread function (PSF) and the local noise. However, to describe an imaging system in a statistical sense, the statistical characteristics of the object are indispensable, and the object is also an integral part of the imaging system. Besides, the directly detected images are not our ultimate goal, but the inputs for calculating statistical features. 

In our statistical model, an imaging system includes four essential elements: an object, an imaging device, noise, and a set of detected images.

\subsection{Object}
\label{ssec:source}
Fundamentally, an object will generate stochastic photons according to quantum theory. Without loss of generality, we consider a one-dimensional discrete object model, where the number of photons generated per unit time at position $k$ can be represented by a random variable $O(k)$. The expectation of $O(k)$ is $\overline{O}(k)$, and the corresponding variance is $\sigma_O^2 (k)$. 

The Mandel Q Parameter \cite{Mandel1979Sub} defined by $Q(k) = \sigma_O^2 (k)/\overline{O}(k)-1$ can indicate the type of the object. If $Q(k)=0$, it refers to a coherent source that obeys Poissonian statistics. A laser is such a kind of source which yields random photon spacing. If $Q(k)<0$, it refers to sub-Poissonian photon statistics, which is a photon number distribution for which the variance is less than the mean. A single atom that emits anti-bunching photons is a typical example. If $Q(k)>0$, it refers to super-Poissonian photon statistics. A thermal light field yields bunched photon spacing, and the variance is larger than the mean. 

Here, we define a new statistic $Z$, which is equal to the variance minus the expectation or mean. For the object region, neglecting its position $k$, we have $Z_O = \sigma_O^2 - \overline{O}$, or equally $Z_O = \overline{O}Q_O$ . The $Z$ quantity can also be applied to the image and the noise, and the corresponding definitions are $Z_I = \sigma_I^2 - \overline{I} = \overline{I}Q_I$ and $Z_N = \sigma_N^2 - \overline{N} = \overline{N}Q_N$. In Section \ref{ssec:zimages}, we will see that $Z_O$, $Z_I$ and $Z_N$ can be linked by a single imaging formula.

\subsection{Imaging Device}
\label{ssec:imagedevice}
%
An imaging device will redistribute the input photons to the image plane, which is controlled by a PSF or more generally a modulation kernel, as mentioned in Equation \ref{eq:convolution} and \ref{eq:modulation} in Section \ref{sec:introduction}. Let's consider a discrete one-dimensional PSF $p(j)$, where $j \in \{ -J, \dots, J \}$ is the pixel position. $p(j)$ is the probability that an input photon from position $0$ on the object plane would fall at position $j$ on the image plane . All $p(j) \ge 0$, and $\sum p(j) = 1$. 

If there are $n$ input photons, and the random variables $X(j)$ indicate the number of photons recorded on pixel $j$, then the vector $\boldsymbol{X} = \{X(-J), \dots, X(J)\}$ follows a multinomial distribution  $f_m(\boldsymbol{x}, n, \boldsymbol{p}) = f_m(x(-J),\dots, x(J), n, p(-J), \dots, p(-J))$ = ${\rm Pr}(X(-J)=x(-J), \cdots, X(J)=x(J))$ whose properties are listed in Table \ref{tab:mnd} \cite{forbes2011statistical}. It is important to show that while $n$ input photons are independent, their outcomes $X(j)$ are dependent because they must be summed to $n$.

\begin{table}[htbp]
\centering
\caption{The properties of a multinomial distribution.}
\begin{tabular}{c|l}
\hline
Property & Description \\
\hline
Trials & $n>0$ (integer) \\
\hline
Probabilities & $p(j), j \in [-J, J]$, $\sum p(j)=1$ \\
\hline
Variables & $x(j) \in \{0, \dots, n \} $, $\sum x(j) = n$ \\
\hline
PMF & $\frac{n!}{x(-J)! \cdots x(J)!}p(-J)^{x(-J)} \cdots p(J)^{x(J)}$\\
\hline
Mean & ${\rm E}(X(j)) = n\, p(j)$ \\
\hline
Variance & ${\rm Var}(X(j)) = n\, p(j)(1-p(j))$ \\
\hline
Covariance & ${\rm Cov}(X(j), X(k)) = -n\, p(j) p(k), \,\, (j \neq k)$ \\
\hline
\end{tabular}
  \label{tab:mnd}
\end{table}

\subsection{Noises}
\label{ssec:noises}
There are several sources of noise in images obtained by CCD or CMOS detectors \cite{blanter2000shot}. The first one is readout noise, which is generated by the on-chip output amplifier. This noise can be reduced to a few electrons with the careful choice of operating conditions. The second one is the dark current, which is caused by thermally generated electrons in the detector and can be minimized by cooling the detector. The third one is photon shot noise, which is a fundamental property of the quantum nature of light.

In this paper, we only treat readout noise and dark current as of the local noise, which mainly originates from detectors and is independent among pixels. The photon shot noise, coming from the object and the imaging device, is actually a signal. 

\subsection{Images}
\label{ssec:images}
A list of images on the same scene with identical parameters, such as gain and exposure time, etc., provide much more information than just one image and are capable of doing statistics. For example, the variance of photon numbers on a pixel indicates the fluctuated level. Furthermore, the covariances between a pair of pixels would uncover the relationship among all pixels in the detector. One image, on the contrary, is not enough for doing such statistics.

We should appropriately set the parameters like gain and exposure time. Otherwise, some information will be lost, and statistics will be meaningless. For instance, small gain and low fluctuation of photon number in a pixel will expunge the statistical properties. If the multiplication of the gain and variation of the photon number is less than one Analog-Digital Unit (ADU), then the actual fluctuation in a corresponding digital image will be invisible.

\section{Statistical Model}
\label{sec:statisticalmodel}

\subsection{Joint Probability Mass Function}
\label{ssec:JPMF}
Random variables can express random trials, and their probability mass function (PMF) in a discrete situation, or probability density function (PDF) in a continuous case, can adequately describe the random trials. If an imaging process is considered as a random trial, then this trial consists of two random sub-trials. First, an object region generates a certain number of photons at position $k$; second, the imaging device projects these photons onto the corresponding detector pixels, which is controlled by the PSF. These two sub-trials are independent of each other, so the PMF of the imaging process is the product of the PMFs of the sub-trials.

Let random variable $O(k)$ represents a segment of an incoherent discrete extended object at position $k, k \in [1, K]$, with PMF of $s(n,k)$, where $n$ denotes the number of photons generated in an unit time. $s(n,k)$ has properties that $\sum_{n=0}^{\infty}s(n,k)=1$ and $\sum_{n=0}^{\infty}n\, s(n,k)=\overline{O}(k)$ where $\overline{O}(k)$ is the expectation of $O(k)$. The $n$ photons at position $k$ will be redistributed into the pixels with positions in $[k-J, k+J]$ of the detector on image plane. Random variables $\boldsymbol{X}(k)$ or $\{X(k-J), \dots, X(k+J)\}$ represent the count rates of photons on these pixels.

The joint PMF of $O(k)$ and $\boldsymbol{X}(k)$ is the product of $s(n,k)$ and $f_m(\boldsymbol{x}(k), n, \boldsymbol{p})$, as described by following expression
\begin{equation}
(O(k), \boldsymbol{X}(k)) \sim s(n,k)f_m(\boldsymbol{x}(k), n, \boldsymbol{p}).
\label{eq:jpmf}
\end{equation}

Let $l=k-j$, $k \in [1, K]$ and $j \in [-J, J]$ and we are interested in the statistical relationships among $X(l)$, $O(k)$ and $p(j)$, which can be deduced theoretically from the joint PMF as listed in Equation \ref{eq:jpmf} and Table \ref{tab:mnd}.

Here, we pay special attention to the expectation, variance, and covariance of
$X(l)$ and connect them to the relevant statistics of a set of observed images. 
In this paper, we only treat the scenario where the source is uncorrelated, which means that the segments of the object are independent of each other.
  
%

\subsection{Mean Image}
\label{ssec:meanofimages}
If we calculated the expectation of $X(k)$ and set it equal to the mean of the images at position $k$, i.e., $E[X(k)] = \overline{I}(k)$,  we got the first fundamental imaging equation
\begin{equation}
\overline{I}(k) = \sum_{j=-J}^{J} \overline{O}(k-j)p(j) + \overline{N}(k),
\label{eq:meanimage}
\end{equation}
where $\overline{O}(k-j)$ was the expectation of the flux density $O(k-j)$ on the object plane, $\overline{N}(k)$ was the mean of the noise $N(k)$ on the image plane. Please see Subsection 6.2 in Appendix for detailed derivation.

We can see that equation \ref{eq:meanimage} is in exactly the same form as equation \ref{eq:convolution}. This relationship is well understood that if we use a unit exposure time, then the mean of a group of images is equivalent to normalizing a long-exposure image.

\subsection{Variance Image}
\label{ssec:varofimages}
If we calculated the variance of $X(k)$,  and set it equal to the variance of the flux of pixel at position $k$, i.e., ${\rm Var}[X(k)] = \sigma_{I}^2(k)$,  we got the following equation
\begin{equation}
\sigma_I^2(k) = \sum_{j=-J}^{J} \left[ \overline{O}(k-j)p(j) +  
          Z_O(k-j)p(j)^2 \right] + \sigma_N^2(k) 
\label{eq:varimage}
\end{equation}
where $Z_O(l)=\overline{O}(l)Q_O(l)=\sigma_O^2(l)-\overline{O}(l), l=k-j$ is the $Z$ quantity of the object, and $Q_O(l)=\sigma_O^2(l)/\overline{O}(l)-1, l=k-j$ is the Mandel Q parameter of the object at position $k-j$, $\sigma_N^2(k)$ is the variance of the noise. The detailed derivation was listed in Subsection 6.3 in Appendix.

A variance image consists of three parts. The first part, since incident photons randomly distribute on the image plane controlled by the PSF,  is the convolution of the expectation of the object and the PSF, which is entirely consistent with the first part of Equation \ref{eq:meanimage}. The second part is unique, it is the convolution of the $Z$ quantity of the object and the square of the PSF. The third part is the contribution of the noise.

\subsection{$Z$ Image}
\label{ssec:zimages}
Subtracting Equation \ref{eq:meanimage} from Equation \ref{eq:varimage}, we get the second fundamental imaging equation
\begin{equation}
Z_I(k) = \sum_{j=-J}^{J}Z_O(k-j)p(j)^2 + 
Z_N(k), 
\label{eq:zimage}
\end{equation}
where $Z_I(k)=\overline{I}(k)Q_I(k)=\sigma_I^2(k)-\overline{I}(k)$ is the $Z$ image, $Z_O(l)=\overline{O}(l)Q_O(l)=\sigma_O^2(l)-\overline{O}(l), l=k-j$ is the $Z$ quantity of the object, and $Z_N(k)=\overline{N}(k)Q_N(k)=\sigma_O^2(k)-\overline{O}(k)$ is the $Z$ quantity of the noise.

Comparing Equation \ref{eq:zimage} with Equation \ref{eq:meanimage}, we find that the $Z$ quantities of the object, noise, and image also satisfy the imaging equation in the form of convolution. The only difference is that the convolution kernel is changed from $p(j)$ to $p(j)^2$. Equation \ref{eq:zimage} leads to a result that the total amount of $Z$ of the object after imaging is not conserved and may be reduced because of $0\le p(j) \le 1$.

The good news is that the resolution of the $Z$ image will increase while reducing the sidelobe level. For example, for a Gaussian PSF, the resolution of the $Z$ image will increase by a factor of $\sqrt{2}$. For an imaging device with a circular aperture, the PSF is an Airy function \cite{airy1835diffraction}, and the corresponding resolution of the $Z$ image can increase by a factor of 1.39.

\subsection{Covariance Image}
\label{ssec:covarofimages}
If we set ${\rm Cov}(X(k), X(l))$ as the covariance between $X(k)$ and $X(l)$, and let it equals to the covariance of a pair of pixels at positions of $k$ and $l$ on image plane, i.e. ${\rm C_I}(k,l) = {\rm Cov}(X(k), X(l))$, we have the third fundamental imaging equation
\begin{equation}
{\rm C_I}(k, l) = \sum_{j=-J}^{J} Z_O(k-j)p(j)p(l-k+j)
\label{eq:covarimage}
\end{equation}
where $Z_O(k-j)$ is the $Z$ quantity of the object. Please see Subsection 6.4 in Appendix for detailed derivation.

If $Z_O(k-j)=Z_O$ is constant around position $k$, then we have 
\begin{equation}
{\rm C_I}(k, l) = {\rm C_I}(l-k) = Z_O \sum_{j=-J}^{J} p(j)p(l-k+j). 
\label{eq:constz}
\end{equation}
For an isolated point object at position $k$, we have
\begin{equation}
{\rm C_I}(k, l) = {\rm C_I}(l-k) = Z_O(k) p(0)p(l-k).
\label{eq:pointz}
\end{equation}
The above two formulas show that after obtaining the covariance image, we can use either a point source or a locally flat extended source to estimate the PSF of the imaging system, and at the same time, calculate the corresponding $Z$ value.

A covariance image is not only free from noise but also provides more information about the object and PSF. In a mean or $Z$ image, each pixel has only one value. In a covariance image, however, each pixel has a curve (one-dimensional imaging) or an image (two-dimensional imaging). This curve or image is closely related to the structure of the object as well as the PSF.

\section{Simulations}
\subsection{Description of the Simulations}
\begin{figure}[htbp]
\centering\includegraphics[width=12cm]{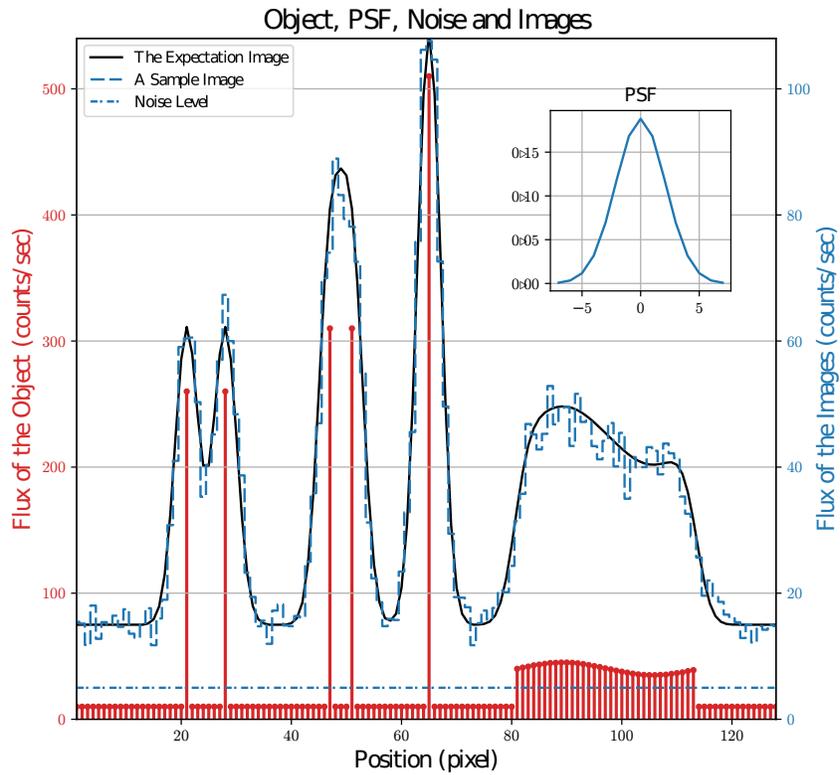}
\caption{The object, PSF, convolution image, and a sample image of the simulations. A set of red vertical lines represent the object whose flux corresponds to the left Y-axis. The PSF is shown in the sub-figure in the upper right corner. A solid black line represents the classic convolution image, a blue step line represents a randomly sampled image, and a blue horizontal dashed line displays the local noise level. The flux of the convolution image, sample image, and noise level correspond to the right Y-axis.}
\label{fig:fig1}
\end{figure}

In the simulations, we use a discrete Gaussian PSF in pixels, with parameter $\sigma$'s value of 3.0 pixel, and the full width at half maximum (FWHM) is about 2.3$\sigma$ = 6.9pix (see the top right sub-figure in Figure \ref{fig:fig1} ). The object, with a length of 128 pixels, is divided into two parts, namely the foreground and the background. The foreground contains a point source, two sets of dual sources with different spacing, and an extended source; the background is constant (see the red vertical line in Figure \ref{fig:fig1}). The flux density of the point source at position 64 is 500 counts/sec. The leftmost dual sources have positions of 21 and 28 respectively, with the spacing of 7 pixels and the flux density of 250 counts/sec. In the classic convolution image, these dual sources can be resolved (see the solid black line in Figure \ref{fig:fig1}). The dual sources further to the central point source have positions of  46pixel and 51pixel, respectively, with the spacing of 5pixel  and the flux density of 300 counts/sec, which is not distinguishable in the convolution image. In the right half of the area, from 81pixel to 113pixel, there is an extended source. Its flux density is a sinusoidal structure with an average of 30 counts/sec and an amplitude of 5 counts/sec. The background has a constant flux density of 10 counts/sec, with Mandel $Q$ parameter of $Q_B=0.2$. The foreground's $Q$ parameters $Q_F$ are set up $-0.2$, $0.0$, and $0.2$.

The steps to obtain a sample image are as follows: First, if the expectation of the flux density of the object at position $k$ is $\overline{O}(k)$,  then we perform a Poisson sampling to obtain a sample with value of $s(k)$, and transform the value of the sample  as follows: $s_a (k) = \alpha (s (k)-\overline{O} (k)) + \overline{O}(k)$, where $\alpha = \sqrt {1 + Q_O}$ and $Q_O$ is the Mandel Q parameter of the object. The random variable $s_a (k)$  represents an object region with the expectation of $\overline{O}(k)$, and the Mandel Q parameter of  $Q_O$. Secondly, $s_a (k)$ photons redistribute into the pixels with position $k-j, j \in [-J, J]$ on the image plane, controlled by a multinomial distribution associated with the PSF. Such dual sampling operations are performed on each object region $O(k)$, and the accumulated photons or signal at pixel $k$ is denoted by $S(k)$. Finally, we add a local noise that is uncorrelated with the signal and has an expectation of 5 counts/sec (see the blue dash-dotted line in Figure \ref{fig:fig1} ) and $Q$ parameter of $Q_N=0.1$.

The statistical results from a list of observed images include the mean image, the variance image, the $Z$ image that represents the difference between the variance and the mean image, and the covariance image. We primarily focus on the $Z$ and covariance images. The $Z$ image can reveal the structures of objects with a non-zero $Q$ parameter and the $Q$ value of the local noise. The covariance image reflects the correlation between a pair of pixels. After obtaining a conventional covariance image, we subtract the mean image from its diagonal values or replace them with the $Z$ image. The corrected covariance image in this way can reveal more information about the object and can separate the background of the object from the local noise.

\subsection{Results}
\begin{figure}[htbp]
\centering\includegraphics[width=12cm]{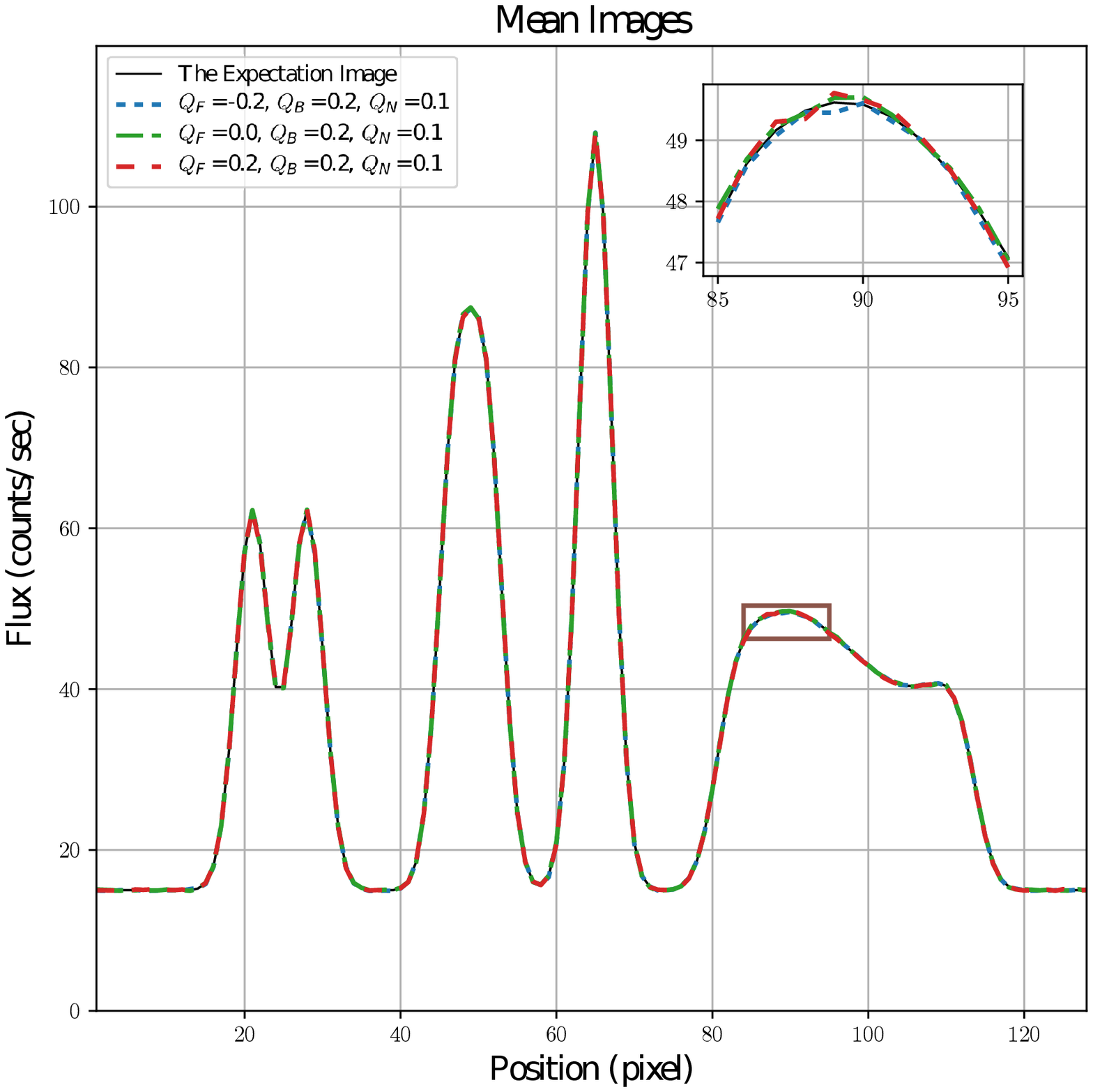}
\caption{Three mean images each calculated from 1000 sample images. The $Q$ parameter of the local noise is $Q_N = 0.1$, and the $Q$ parameter of the background in the object is $Q_B = 0.2$. Regarding the $Q$ parameter of the foreground in the object, we set three different values, i.e., $Q_F$ is -0.2, 0.0, or 0.2, corresponding to the blue, green, and red dashed lines in the figure respectively. Also, the convolution image is shown in a solid black line.}
\label{fig:fig2}
\end{figure}

Figure \ref{fig:fig2} shows three mean images, as well as the convolution image for comparison. The results verify the inference in section \ref{ssec:meanofimages} that a mean image approaches the classic convolution image. The difference between them (see the zoomed-in area in the upper right sub-figure), due to a limited number of samples,  becomes smaller as the number of sample images increases. Also, the results show that there is no significant difference between mean images, even if the object and the local noise have different $Q$ parameters.

\begin{figure}[htbp]
\centering\includegraphics[width=12cm]{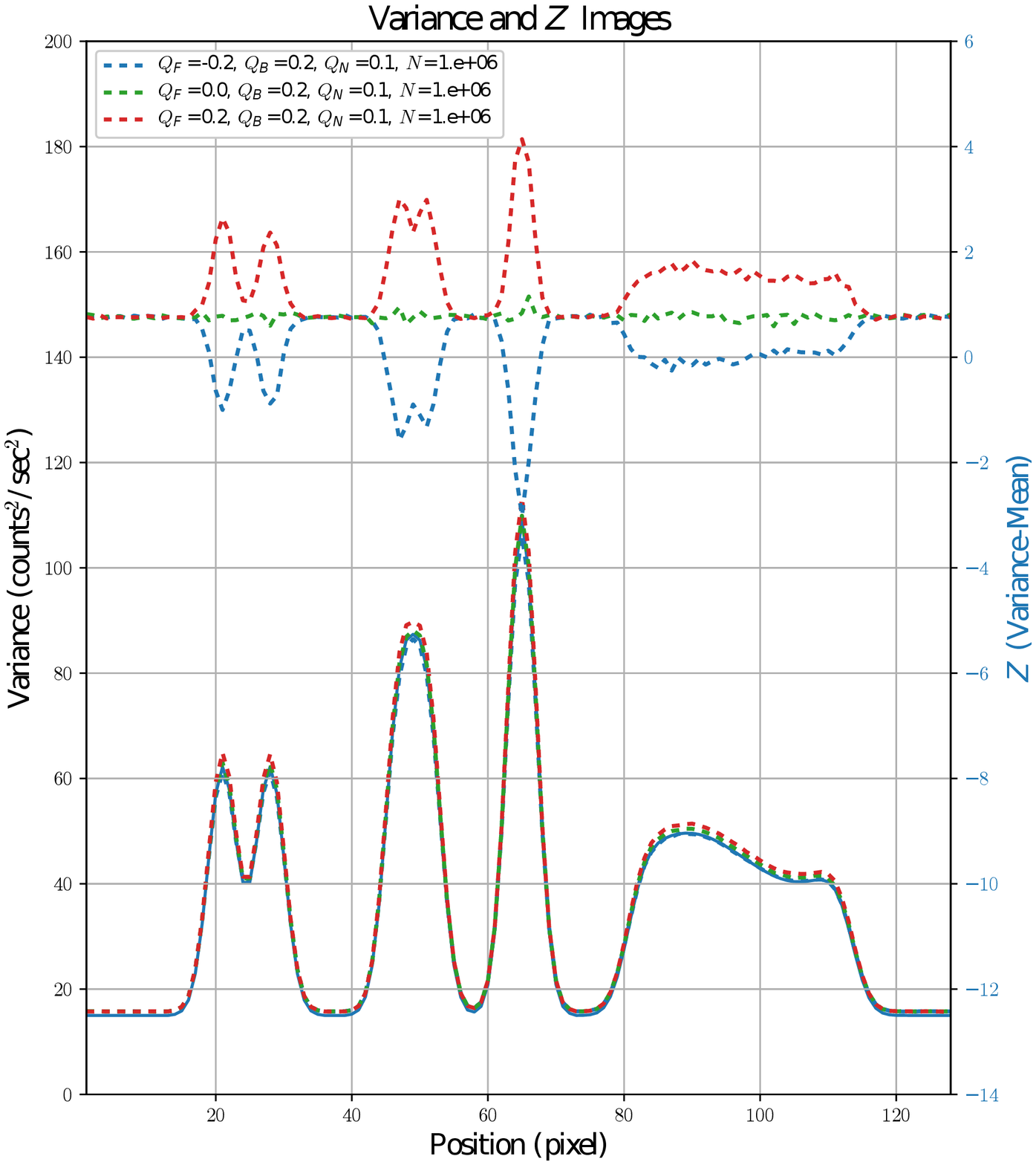}
\caption{A set of variance and $Z$ images. The number of sample images is $N$=1.0e+6. The $Q$ parameters of the local noise and the background in the object are $Q_N = 0.1$ and $Q_B = 0.2$ respectively. Regarding the $Q$ parameter of the foreground in the object, we set three different values, i.e., $Q_F$ is -0.2, 0.0, or 0.2, corresponding to the blue, green, and red dashed lines in the figure. The amplitudes of the variance images (bottom) correspond to the scale of the left Y-axis, and the amplitudes of the $Z$ images (top) correspond to the scale of the right Y-axis. }
\label{fig:fig3}
\end{figure}

Figure \ref{fig:fig3} shows a set of variance images and corresponding $Z$ images. We can see that the variance and mean images are roughly similar in structure. However, when the $Q$ parameters of the object and noise take non-zero values, there will be a significant difference between the variance image and the mean image. The $Z$ image (the upper part of Figure \ref{fig:fig3}), obtained by subtracting the mean image from the variance image, can reveal the structure of the object and the characteristics of the noise. As pointed out by Equation 5, the resolution of the $Z$ image is indeed higher than that of the mean image. For example, the dual-source structure at positions 46 and 51 cannot be resolved in the mean image but can be clearly resolved in the $Z$ image. Combined with the mean image, the $Z$ image can also be used to calculate the $Q$ parameter of the object quantitatively. However, in the $Z$ image, the background part of the object is still coupled with the local noise and cannot be distinguished.

\begin{figure}[htbp]
\centering{
\includegraphics[width=7cm]{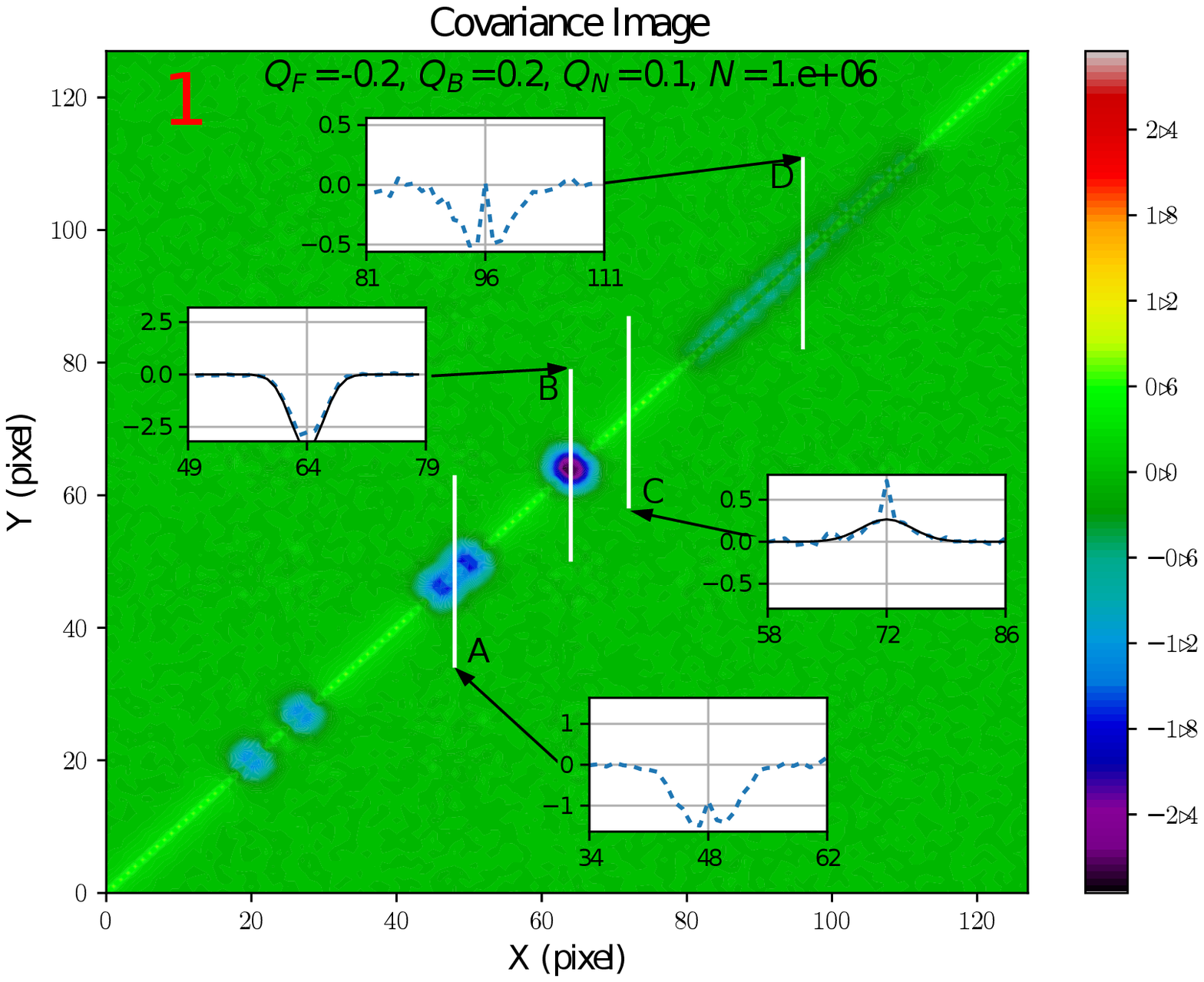}\includegraphics[width=7cm]{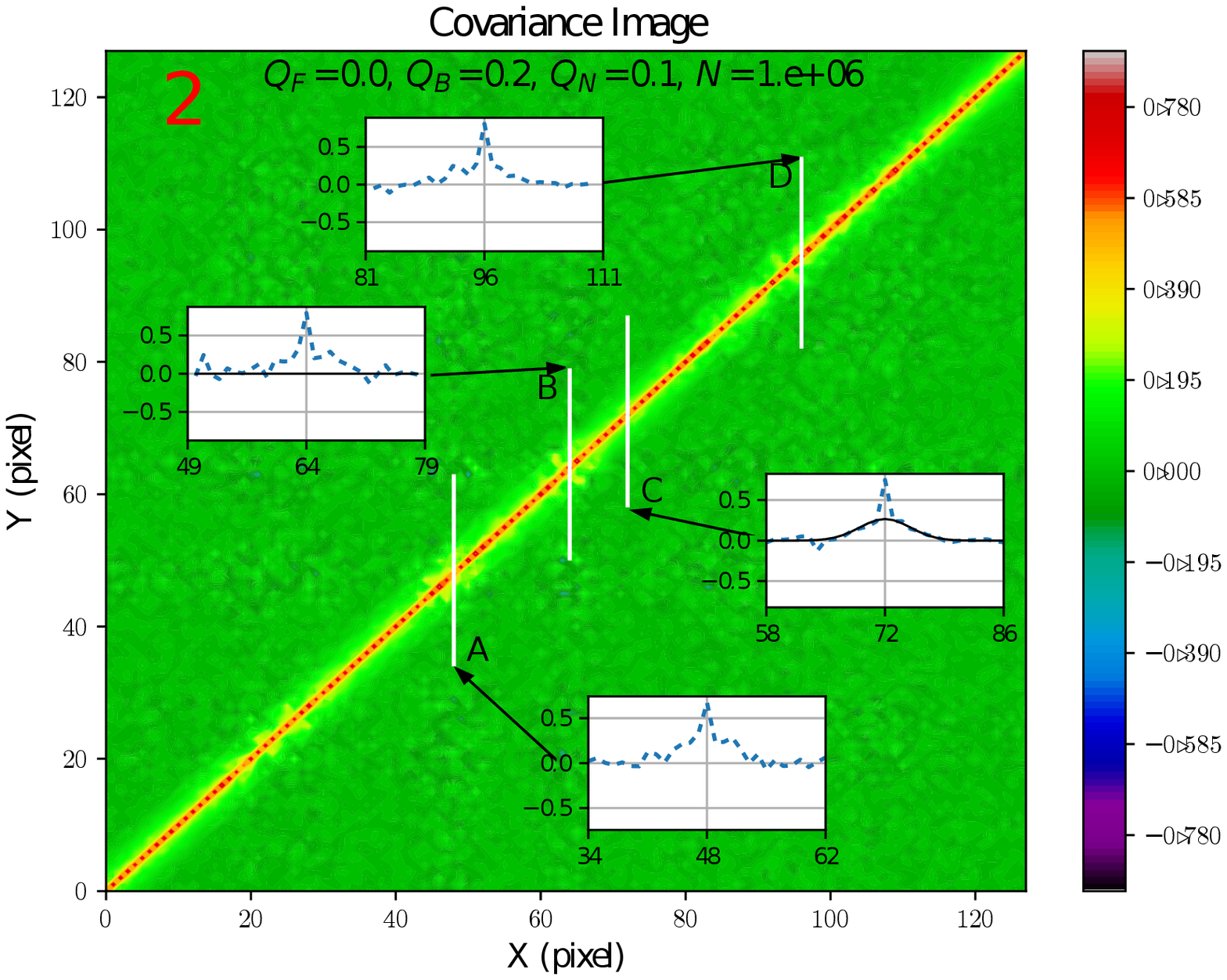}
\includegraphics[width=7cm]{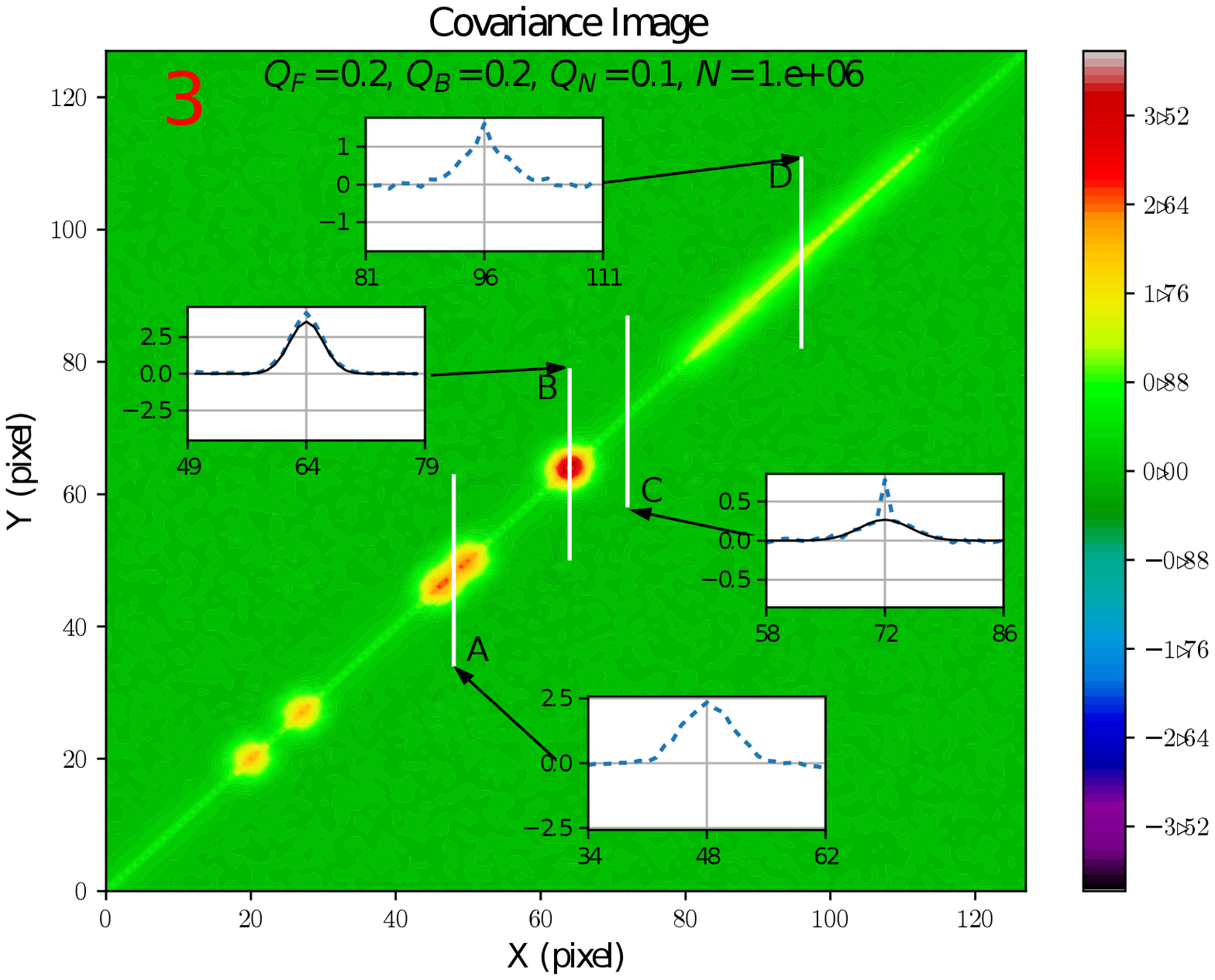}
}
\caption{A set of covariance images. The number of sample images is $N=1.0\mathrm{e}+6$. The $Q$ parameters of the local noise and the background of the object are $Q_N=0.1$ and $Q_B=0.2$ respectively. We selected three different foreground $Q$ parameters: i.e. $Q_F=-0.2$ (Map 1),  $Q_F=0.0$ (Map 2) and $Q_F=0.2$(Map 3). We also chose 4 slices at the positions of 48, 72, 64, and 96 in each covariance image, labeled as {\bf A}, {\bf B}, {\bf C}, and {\bf D} and their covariance curves are shown in the corresponding 4 sub-figures.}
\label{fig:fig4}
\end{figure}

The corrected covariance images, as shown in Figure \ref{fig:fig4}, reveal more information about the objects. Since the local noise does not correlate with each other, it appears as a thin line on the diagonal on the covariance image. However, both foreground and background are modulated by the PSF, which appears as a two-dimensional extended structure on the covariance image. Therefore, the structures in a covariance image directly reflect the $Z$-distribution of the object. Combined with the mean image in Figure \ref{fig:fig2}, the distribution of $Q$ parameters of the object can be further uncovered, and thus the physical characteristics of the object can be explored.

The corrected covariance image can be used to estimate the $Z$ quantities of the locally invariable structure in the object and the local noise. First, find a position where the structure remains constant locally on the covariance image, for example, like the slice {\bf C} in Figure \ref{fig:fig4}, and extract the corresponding covariance curve. The central point of the covariance curve is the $Z$ quantity at that position, represented by $Z(k)$, which may come from either the background of the object or the local noise. Without this central point, Equation \ref{eq:constz} can describe the rest of the curve related only to the invariable structure. Through fitting, we can estimate the $Z$ quantity ($Z_B$) of the background. After that, we subtract the $Z_B$ from the $Z(k)$ and estimate the $Z$ value of the noise, i.e., $Z_N$. If the $Q$ parameter or the mean value $\overline{N}$ of the local noise is known, then we can calculate the $\overline{O}_B$ and $Q_B$ of the background at the same time, by combining Equation \ref{eq:meanimage}.

The same method can be applied to point sources, such as slice B in Figure \ref{fig:fig4}. Therefore, in this statistical model, we can directly separate the locally constant background or point sources of the object from the local noise with the help of the $Z$ image and covariance image, although they are mixed in the mean image.

\section{Discussion}
In the previous derivation and simulation, we only considered uncorrelated objects. Although most physical objects are uncorrelated, there are also cases of correlated or partially correlated objects. For correlated objects, the fundamental joint probability mass function, shown in Equation \ref{eq:jpmf}, still applies, and Equation \ref{eq:meanimage} describing the mean image is still applicable. However, the corresponding formulas describing the variance image, the $Z$ image, and the covariance image need to be re-derived. We can also use numerical simulations to study statistical properties when imaging correlated objects.

In future work, we will also extend the statistical model to the general modulation imaging process where the PSF is no longer shift-invariant, as described in Equation \ref{eq:modulation}, to see what kind of formulas would be suitable for the description. Similarly, numerical simulation will also be a potent tool to explore the statistical properties of the model.

The intrinsic resolution of the $Z$ image is higher than that of the mean image. If we exploit the information provided by the covariance image, the resolution of the reconstructed object can be further improved. How to use Equation \ref{eq:covarimage} for such reconstruction will be one of the main objectives in the future.

The photon bunching effects only occur when the time resolution is less than the coherence time, which is usually in the order of picoseconds in the optical band. Therefore, we need high-speed cameras with a time resolution of picoseconds. Assmann et al. redesigned a streak camera and used it to measure the high-order photon bunching effects \cite{Assmann2009Higher}. In addition, a two-dimensional Electron-Bombarded CCD (EBCCD) \cite{Bryukhnevich1992Picosecond, Hirvonen2014Photon, Hirvonen2015Photon}readout device can be used in a picosecond electron-optical information system, which will be very useful for statistical imaging systems.

In radio or microwave band, a receiver can simultaneously record the amplitude and phase of the electromagnetic field. The received data can be used not only as an interferometer but also as a radiometer, which will measure the visibility or flux density of an object, respectively. Lieu \cite{Lieu2017Improvement} pointed out that the photon bunching effect would be used to improve the accuracy of a radiometer. With the help of the results of this statistical model, imaging systems at the radio band will provide new methods to estimate and mitigate local noises and interferences and pick up some rare objects in complex scenes.

In many cases, the intrinsic coherence time of an object is short, maybe at the attoseconds scale for X-rays objects. However, some astronomical objects, such as pulsars, have coherent macroscopic variations in the order of milliseconds to seconds, from low radio band to high $\gamma$-ray band. The statistical model proposed in this paper is also applicable to such macroscopic coherent changes and is used to analyze the $Z$ quantities and $Q$ parameters of the objects as well as local noises.

Combining the $Z$ image, covariance image, and mean image, we could reconstruct the $Z$-distribution and $Q$-distribution of the object, which are closely related to the physical properties of the object. Consequently, this statistical model opens a new realm for imaging systems to explore the physical world.

\section{Conclusions}
We established a statistical model for imaging systems and obtained three fundamental imaging formulas. Among them, the first formula is entirely consistent with the classic convolution equation, while the other two formulas reveal new laws. The $Z$ quantity of the object, image, and noise can also be linked with an elegant convolution equation. Also, the $Z$ quantity of the object and the covariance image satisfy the third imaging equation. So, besides the flux density, the $Z$ quantity of an object is also imageable, which opens a new window for imaging systems. We believe that the statistical model proposed in this paper is generally applicable, from low-frequency radio band to high-energy Gamma-ray band, from such as physics, biology to astronomy.

\section*{Funding Information}
National Key R\&D Program of China (2016YFA0400802); National Natural Science Foundation of China (NSFC) (11373025).

\section*{Acknowledgments}
The author would like to thank Mr. Wei Dou for his valuable comments.

\bibliography{OSA-template}

\section*{Appendix}
\subsection{Imaging Model}
\label{app:imgmodel}
For a pixel $k$ in the image plane, its recorded photons $I(k)$ consists of the signals $S(k)$ come from the object and the local noise $N(k)$, so
\begin{equation}
    I(k) = S(k) + N(k),
\end{equation}
where $I(k)$, $S(k)$ and $N(k)$ are all random variables.

The signals $S(k)$ may come from different areas of the object. Let $X_j(k)$ denote the signal comes from the object region $O(k-j)$. Altogether, there are $2J+1$ regions that can influence the pixel $k$ of the image, so the total signals are 
\begin{equation}
    S(k) = \sum_{j=-J}^{J} X_j(k),
\end{equation}
where $X_j(k)$ are random variables and are independent of each other. 

The variable $X_j(k)$ is associated with the object region $O(k-j)$ which has a probability mass function (PMF) of $s(n,k-j)$ , with the expectation of
\begin{eqnarray}
{\rm E}[O(k-j)] &=& \sum_{n=0}^{\infty}n\, s(n,k-j) \\
&= & \overline{O}(k-j), \label{eq:expobj}
\end{eqnarray}
and variance of
\begin{eqnarray}
{\rm Var}[O(k-j)] &=& {\rm E}[O^2(k-j)] - {\rm E}[O(k-j)]^2 \\ 
&= & \sigma_O^2(k-j), \label{eq:varobj}
\end{eqnarray}
where
\begin{equation}
    {\rm E}[O^2(k-j)] = \sum_{n=0}^{\infty}n^2 s(n,k-j).
\end{equation}

The object region $O(k-j)$ is coupled with a set of random variables $\{ X(k-j+i) \}, i \in [-J,J]$ or $\{ X(k-j-J),\dots, X(k-j+J) \}$ or $\boldsymbol{X}(k-j)$ which represent the photons recorded in pixels $k-j+i$ and have a PMF of multinomial distribution $f_m(\boldsymbol{x}(k-j), n, \boldsymbol{p})$ \cite{forbes2011statistical}. Since we here only care about the photons in  pixel $k$, therefore $i=j$ and $X_j(k)=X(k)$ for the object region $O(k-j)$, with the following properties
\begin{eqnarray}
{\rm E}^m[X(k)] &=& n\, p(j) \label{eq:expm}\\ 
{\rm Var}^m[X(k)] &=& n\, p(j)(1-p(j)) \label{eq:varm}\\
{\rm Cov}^m[X(k), X(k+l)] &=& -n\, p(j) p(j+l),
\end{eqnarray}
where $j,j+l \in [-J,J]$.

As indicated in Section 3.A in the main article, the random variables $O(k-j)$ and $\boldsymbol{X}(k-j)$ have a joint PMF of
\begin{equation}
    f_{\rm join}(O(k-j),\boldsymbol{X}(k-j)) = 
s(n,k-j)f_m(\boldsymbol{x}(k-j), n, \boldsymbol{p}). 
\end{equation}
This joint PMF is the core of the imaging model, through which we can obtain the statistical characteristics of a mean image, a variance image and a covariance image.

\subsection{Mean Image}
\label{app:meanimage}
A mean image is the average of a set of images with isochronous exposure. Besides, based on the joint PMF, we can derive an expression of the expectation image, that is, the expectation of the observed image. Statistically speaking, the mean image is an estimate of the expectation image. With the mean image, the structures of the object can be inferred based on the derived expression.

Firstly, we calculate the expectation of $X_j(k)$ which means the photons from object region $O(k-j)$ recorded in pixel $k$, 
\begin{eqnarray}
    {\rm E}[X_j(k)] &=& \sum_{n=0}^{\infty}          
                   s(n,k-j) {\rm E}^m[X(k)] \\
    & = & \sum_{n=0}^{\infty}          
                   s(n,k-j)n\, p(j) \\
    & = & \overline{O}(k-j)p(j),
\end{eqnarray}
where we set $X_j(k) = X(k)$ for the object region $O(k-j)$.
Then, the expectation of the total photons recorded in pixel $k$ is  
\begin{equation}
    {\rm E}[S(k)] = \sum_{j=-J}^{J} {\rm E}[X_j(k)] = \sum_{j=-J}^{J}\overline{O}(k-j)p(j).
\end{equation}
Finally, we obtain the expression of a mean image $\overline{I}(k)$,
\begin{equation}
    \overline{I}(k) = {\rm E}[I(k)] = \sum_{j=-J}^{J}\overline{O}(k-j)p(j) + \overline{N}(k),
\end{equation}
where $\overline{N}(k)$ is the mean of the noise.

\subsection{Variance Image}
\label{app:varimage}
For a random variable $S(k)$, we have
\begin{equation}
\label{eq:varsk}
    {\rm Var}[S(k)] = {\rm E}[S(k)^2] - {\rm E}[S(k)]^2.
\end{equation}
Firstly, we calculate the first part of Equation \ref{eq:varsk}, 
\begin{eqnarray}
    {\rm E}[S^2(k)] &=& {\rm E}[\sum_{j=-J}^{J}X_j(k) \sum_{i=-J}^{J}X_i(k)] \\
    &=& \sum_{j=-J}^{J} \sum_{i=-J}^{J} {\rm E}[X_j(k)X_i(k)]\\
    &=& \sum_{j=-J}^{J}{\rm E}[X_j^2(k)] + \sum_{i,j=-J,i\neq j}^{J}{\rm E}[X_j(k)X_i(k)] \label{eq:expectsk}
\end{eqnarray}
Again, from Equation \ref{eq:expectsk}, we can see that the 
summation was cut into two parts. Part one is 
\begin{eqnarray}
    {\rm E}[X_j^2(k)] &=& \sum_{n=0}^{\infty}          
                   s(n,k-j) {\rm E}^m[X^2(k)] \\
    & = & \sum_{n=0}^{\infty}s(n,k-j)({\rm Var}^m[X(k)] + {\rm E}^m[X(k)]^2)\\
    & = & \sum_{n=0}^{\infty}s(n,k-j)\\
    &   & \left( n\,p(j)(1-p(j)) + n^2p(j)^2 \right) \\
    & = & \overline{O}(k-j)p(j)(1-p(j)) + \\ 
    &   & \left(\sigma_O^2(k-j)+\overline{O}(k-j)^2 \right) p(j)^2,
\end{eqnarray}
where the results in Equation \ref{eq:expobj}, \ref{eq:varobj}, \ref{eq:expm}, and \ref{eq:varm} were used. Next, we calculate part two in Equation \ref{eq:expectsk}. Since $X_j(k)$ and $X_i(k), i\neq j$ represent the photons come from different object region $O(k-j)$ and $O(k-i)$, they are independent. Therefore, we have
\begin{eqnarray}
    {\rm E}[X_j(k)X_i(k)] &=& {\rm E}[X_j(k)]{\rm E}[X_i(k)]\\
    &=& \overline{O}(k-j)p(j) \overline{O}(k-i)p(i).
\end{eqnarray}
Then, 
\begin{eqnarray}
    {\rm E}[S^2(k)] &=& \sum_{j=-J}^{J}\overline{O}(k-j)p(j)(1-p(j)) + \\ 
    & & \sum_{j=-J}^{J}\left(\sigma_O^2(k-j)+\overline{O}(k-j)^2 \right) p(j)^2 \\
    & & + \sum_{i,j=-J,i\neq j}^{J}\overline{O}(k-j)\overline{O}(k-i)p(j)p(i).
\end{eqnarray}

Now, let's calculate the second part of Equation \ref{eq:varsk},
\begin{eqnarray}
    {\rm E}[S(k)]^2 &=& \left[\sum_{j=-J}^{J}\overline{O}(k-j)p(j) \right]^2 \\ 
    &=& \sum_{j=-J}^{J}\overline{O}(k-j)^2p(j)^2 \\
    & & + \sum_{i,j=-J,i\neq j}^{J}\overline{O}(k-j)\overline{O}(k-i)p(j)p(i).
\end{eqnarray}
Substituting the results of ${\rm E}[S^2(k)]$ and ${\rm E}[S(k)]^2$ into Equation \ref{eq:varsk}, we obtain
\begin{eqnarray}
    {\rm Var}[S(k)] &=& \sum_{j=-J}^{J}\overline{O}(k-j)p(j)(1-p(j)) + \\ 
    & & \sum_{j=-J}^{J}\sigma_O^2(k-j)p(j)^2 \\
    &=& \sum_{j=-J}^{J}\overline{O}(m)p(j) + \overline{O}(m)Q_O(m)p(j)^2\\
    &=& \sum_{j=-J}^{J}\overline{O}(m)p(j) + Z_O(m)p(j)^2\\
\end{eqnarray}
where $Q(m) = \sigma_O^2(m)/\overline{O}(m) - 1$ is the Mandel $Q$ parameter of the object, $Z_O(m)=\overline{O}(m)Q_O(m) = \sigma_O^2(m)-\overline{O}(m)$ is the $Z$ quantity of the object and $m=k-j$.

Since the signals and the noise are independent of each other, we finally get the expressions of a variance image $\sigma_I^2(k)$
\begin{equation}
\sigma_I^2(k) = \sum_{j=-J}^{J} \left[ \overline{O}(k-j)p(j) +  
          Z_O(k-j)p(j)^2 \right] + \sigma_N^2(k), 
\end{equation}
where $\sigma_N^2(k)$ is the variance of the noises.

\subsection{Covariance Image}
\label{app:covimage}
Firstly, we consider the covariance of two signals $S(k)$ and $S(l)$ at pixels $k$ and $l$ respectively. We have
\begin{eqnarray}
    {\rm Cov}[S(k),S(l)] &=& {\rm E}[S(k)S(l)] - {\rm E}[S(k)]{\rm E}[S(l)]\\
    &=& \sum_{j=-J}^{J}\sum_{i=-J}^{J} [{\rm E}[X_j(k)X_i(l)] - \\
    & & {\rm E}[X_j(k)]{\rm E}[X_i(l)]]
\end{eqnarray}
In the above equations, if $k-j\neq l-j$, then $X_j(k)$ and $X_i(l)$ represent the recorded photons come from different regions of the object, and they are independent of each other. In this case, the covariance between $X_j(k)$ and $X_i(l)$ is zero. 

Consequently, we only need to consider the case 
where $k-j=l-i$, i.e. $X_j(k)$ and $X_i(l)$ represent the photons come from the same source region $O(k-j)$. We have
\begin{eqnarray}
    {\rm E}[X_j(k)X_{i}(l)] &=& {\rm E}[X(k)X(l)] \\
    &=& \sum_{n=0}^{\infty}s(n,k-j){\rm E}^m[X(k)X(l)] \\
    &=& \sum_{n=0}^{\infty}s(n,k-j)[{\rm Cov}^m[X(k),X(l)] \\
    & & +  {\rm E}^m[X(k)]{\rm E}^m[X(l)] ] \\
    &=& \sum_{n=0}^{\infty}s(n,k-j)(-n\,p(j)p(l-k+j) \\
    & & + n^2p(j)p(l-k+j)) \\
    &=& (-\overline{O}(k-j) + \sigma_O^2(k-j) + \\
    & & \overline{O}(k-j)^2)p(j)p(l-k+j),
\end{eqnarray}
and
\begin{eqnarray}
    {\rm E}[X_j(k)]{\rm E}[X_i(l)] &=& {\rm E}[X(k)]{\rm E}[X(l)] \\
    &=& \overline{O}(k-j)^2p(j)p(l-k+j). \\
\end{eqnarray}

Considering the fact that the signals $S(k)$ and noises $N(k)$ are independent of each other, the covariance between them is zero, so we finally get the expression of a covariance image ${\rm C}_I(k,l)$
\begin{eqnarray}
    {\rm C}_I(k,l) &=& {\rm C}_I(I(k),I(l)) \\
    &=& {\rm Cov}[S(k),S(l)] \\ 
    &=& \sum_{j=-J}^{J}(\sigma_O^2(k-j)-\overline{O}(k-j))\\
    & & p(j)p(l-k+j)\\
    &=& \sum_{j=-J}^{J}\overline{O}(m)Q_O(m)p(j)p(l-m) \\
    &=& \sum_{j=-J}^{J} Z_O(m)p(j)p(l-m),
\end{eqnarray}
where $Q_O(m) = \sigma_O^2(m)/\overline{O}(m)-1$ is the Mandel $Q$ parameter of the object, $Z_O(m) = \overline{O}(m)Q_O(m) = \sigma_O^2(m)-\overline{O}(m)$ is the $Z$ quantity of the object and  $m=k-j$.

\end{document}